\documentclass[aps,prl,twocolumn,showpacs,superscriptaddress,groupedaddress]{revtex4-1}  

\usepackage{amssymb}
\usepackage{amsthm}
\usepackage{amsmath}
\usepackage{graphicx,color}
\usepackage{acronym}
\usepackage{amsfonts}
\usepackage{amscd}
\usepackage{dcolumn}   
\usepackage{bm}        
\usepackage{braket}

\usepackage[english]{babel}

\renewcommand{\qed}{\bull \medskip}
\usepackage[colorlinks=true]{hyperref}

\newtheorem{theorem}{Theorem}[section]

\newtheorem{algo}{Algorithm}[section]
\newtheorem{example}{Example}[section]

\def\qed{\hfill {$ \Box $} \medskip}

\begin{document}

\title{\bf Separability discrimination and decomposition of\\ $m$-partite quantum mixed states}

\author{Ying Li}
\affiliation{Department of Mathematics, National University of
Defense Technology, Changsha, Hunan 410073, China.}

\author{Guyan Ni}\thanks{Corresponding author.}
\affiliation{Department of Mathematics, National University of
Defense Technology, Changsha, Hunan 410073, China.}


\begin{abstract}
We propose an $E$-truncated $K$-moment and semidefinite relaxations (ETKM-SDR) method to check whether an $m$-partite quantum mixed state is separable or not and give a decomposition for it if it is. We first convert the separability discrimination problem of mixed states to the positive Hermitian decomposition problem of Hermitian tensors. Then, employing the $E$-truncated $K$-moment method, we obtain an optimization model for discriminating separability. Moreover, applying semidefinite relaxation method, we get a hierarchy of semidefinite relaxation optimization models and propose an algorithm for detecting  the separability of mixed states. The algorithm can also be used for symmetric and non-symmetric decomposition of separable mixed states. By numerical examples, we find that not all symmetric separable states have symmetric decompositions.

\medskip


\end{abstract}


\maketitle


Quantum entanglement was first introduced by Einstein, Podolsky, and Rosen \cite{Ein35} and Schir\"{o}dinger \cite{Sch35}. Entanglement plays a central role in applications of quantum information science as well as in the foundations of quantum theory.
Hence, the question of whether a given state is entangled or separable is one of the fundamental problems in quantum information theory (cf. G\"{u}hne, \cite{Guh2004}).
There are some elegant methods for the separability checking problem have been derived, such as Bell inequality \cite{Bell1964}, positive partial transposition criterion \cite{Per1996,HHH1996}, computable cross norm or realignment criterion (CCNR criterion) \cite{ZHS1998,VW2002}, covariance matrix criterion \cite{HT2003,Guh2004,GHG2007}, correlation matrix criterion \cite{Vic2007}, entanglement witness \cite{HHH1996,Ter2000,LKC2000}, and other methods. The separability detecting problem is a longer standing problem and has attracted great interest in the last 20 years \cite{LWFL2014,OH2010,Bern2015,XZZ2016,SLL2018}. All these methods are based on the sufficient or necessary conditions for entanglement.

In this letter, we use the tensor optimization method to detect a given $m$-partite mixed state it is separable or not. If it is not, we get a certificate for that. If it is, we obtain a decomposition of the state.
In the process, we first take an Hermitian tensor to represent a mixed state, and substitute the separability discrimination problem of mixed states by the positive Hermitian decomposition problem of Hermitian tensors.
And then, employing the $E$-truncated $K$-moment method, we show that an $m$-partite quantum mixed state is separable if and only if there exists an atomic Borel $K$-measures $\mu$ such that the condition (CD1) is satisfied. Then we convert the separability discrimination problem to a moment optimization problem. Finally, applying the semidefinite relaxations method, we propose an algorithm for detecting a mixed state if it is separable or not and decomposing the state if it is. There are similar results for symmetric separability discrimination. Theoretically, by using the algorithm, we can check the separability or symmetric separability of any mixed state and decompose it if it is separable.


{\it Notations.}---The symbol $\mathbb{N}$ (resp., $\mathbb{R}$, $\mathbb{C}$) denotes the set of nonnegative integers (resp., real numbers, complex numbers). For every $k\in \mathbb{N}$, denote $[k] := \{1, \cdots , k\}$.
The symbol $\mathbb{R}[x]: = \mathbb{R}[x_1, \cdots, x_n]$ denotes the ring of polynomials in $x := (x_1, \cdots , x_n)$ with real coefficients.
For $\alpha\in \mathbb{N}^n$, denote $|\alpha|=\alpha_1+\cdots+\alpha_n$, $\mathbb{N}^n_d := \{\alpha \in \mathbb{N}^n : |\alpha| \leq d\}$.
For $x \in \mathbb{R}^n$ and $\alpha\in \mathbb{N}^n$, denote $x^\alpha:=x_1^{\alpha_1}x_2^{\alpha_2}\cdots x_n^{\alpha_n} $. Denote by $[x]_d: = (x^\alpha)_{\alpha\in \mathbb{N}^n_d}$ the vector of monomials, whose exponents are from $\mathbb{N}^n_d$.
Similarly, denote by $\mathbb{R}[x]_d :=\{\sum_{\alpha} p_\alpha x^\alpha : p_\alpha\in \mathbb{R}, \alpha\in \mathbb{N}^n_d \}$ the ring of polynomials with degree no more than $d$.


{\it Complex tensor representation of quantum states.}---An {\it $m$th-order complex tensor} denoted by $\mathcal{A}:=(\mathcal{A}_{i_1...i_m})\in \mathbb{C}^{n_1\times\cdot\cdot\cdot\times n_m}$ is a multiway array consisting of numbers $\mathcal{A}_{i_1...i_m}\in\mathbb{C}$ for all $i_k\in [n_k]$ and $k\in [m]$. A square tensor $\mathcal{S}=(\mathcal{S}_{i_1...i_m})\in \mathbb{C}^{n\times\cdot\cdot\cdot\times n}$ is called symmetric if its entries $\mathcal{S}_{i_1...i_m}$ are invariant under any permutation of $[i_1,...,i_m]$. Given $m$ vectors $\mathbf{z}^{(i)}\in \mathbb{C}^{n_i},\ i\in [m] $, a rank-1 complex tensor $\otimes_{i=1}^m \mathbf{z}^{(i)} $ is defined as
$$(\otimes_{i=1}^m \mathbf{z}^{(i)})_{i_1...i_m}:=z^{(1)}_{i_1}\cdot \cdot \cdot z^{(m)}_{i_m}.$$

An $m$-partite pure state $|\psi\rangle$ of a composite quantum system can be regarded as a normalized element in a Hilbert space $\otimes_{k=1}^{m} \mathbb{C}^{n_k}$. Assume that $\{|e^{(k)}_{i_k}\rangle: i_k\in [n_k] \}$ is an orthonormal basis of $\mathbb{C}^{n_k}$. Then $\{|e^{(1)}_{i_1} e^{(2)}_{i_2} \cdots e^{(m)}_{i_m}\rangle: i_k\in [n_k];\ k\in [m] \}$ is an orthonormal basis of $\otimes_{k=1}^{m} \mathbb{C}^{n_k}$. Hence, $|\psi\rangle$ can be written as
\begin{equation}\label{psi}
|\psi\rangle :=\sum_{i_1,\cdots,i_m=1}^{n_1,\cdots,n_m} x_{i_1\cdots i_m} |e^{(1)}_{i_1} e^{(2)}_{i_2} \cdots e^{(m)}_{i_m}\rangle,
\end{equation}
where $x_{i_1...i_m}\in\mathbb{C}$. Denote $\chi_{|\psi\rangle}:=(x_{i_1\cdots i_m}) $. Then $\chi_{|\psi\rangle} $ is called a {\it corresponding tensor} of $|\psi\rangle$ under the orthonormal basis.
$|\psi\rangle$ is called symmetric if these amplitudes are invariant under permutations of the parties, which is corresponding to a symmetric complex tensor. A separable $m$-partite pure state is denoted as $|\phi\rangle :=\otimes_{k=1}^m |\phi^{(k)}\rangle$. It is clear that the separable pure state is  corresponding to a rank-one complex tensor $\otimes_{k=1}^m v^{(k)} $, where $v^{(k)}\in \mathbb{C}^{n_k} $, more details please see \cite{NQB14}.

Hermitian tensor is an extension of Hermitian matrices.
A $2m$th-order tensor $\mathcal{H}=(\mathcal{H}_{i_1...i_m j_1...j_m})\in \mathbb{C}^{n_1\times\cdots\times n_m\times n_1\times\cdots \times n_m} $ is called a {\it Hermitian tensor} if $\mathcal{H}_{i_1...i_m j_1...j_m}=\mathcal{H}^*_{j_1...j_m i_1...i_m}$ for every $i_1, ..., i_m$ and  $j_1, ..., j_m$, where $x^*$ denotes the complex conjugate of $x$. A Hermitian tensor $\mathcal{H}$ is called a {\it symmetric Hermitian tensor} if $ n_1=\cdots=n_m $ and its entries $\mathcal{H}_{i_1...i_m j_1...j_m}$ are invariant under any permutation operator $P$ of $\{1,...,m\}$, i.e.,
$\mathcal{H}_{i_1...i_m j_1...j_m}=\mathcal{H}_{P[i_1...i_m] P[j_1...j_m]}.$
The space of all Hermitian tensors $\mathcal{H} \in \mathbb{C}^{n_1\times\cdots\times n_m\times n_1\times\cdots \times n_m}$ is denoted by $\mathbb{H} {[n_1, \ldots, n_m]}$, and the space of all  symmetric Hermitian tensors is denoted by $s\mathbb{H}[m,n]$ for convenience, respectively.

For a Hermitian tensor
$\mathcal{H}\in \mathbb{H} {[n_1, \ldots, n_m]}$,
if it can be written as
\begin{equation} \label{Eq:rank-oneHermitDecomp}
\mathcal{H}=\sum_{i=1}^r \lambda_i \,  u_i^{(1)}\otimes \ldots \otimes u_i^{(m)} \otimes u_i^{(1)*}\otimes \ldots \otimes u_i^{(m)*}
\end{equation}
for $\lambda_i\in \mathbb{R}$, $u_i^{(j)}\in \mathbb{C}^{n_j}$
and $\|u_i^{(j)} \|=1$,
then $\mathcal{H}$ is called {\it Hermitian decomposable}.
In this case, (\ref{Eq:rank-oneHermitDecomp})
is called a {\it Hermitian decomposition} of $\mathcal{H}$.
If all $\lambda_i>0$, then (\ref{Eq:rank-oneHermitDecomp}) is called a {\it positive Hermitian decomposition} of $\mathcal{H}$, and $\mathcal{H}$ is called {\it positive Hermitian decomposable}.

Similarly, for a quantum mixed state $\rho$, its density matrix is always written as
\begin{equation}\label{eq:mixedstate}
  \rho=\sum_{i=1}^k p_i |\psi_i \rangle \langle \psi_i |,
\end{equation}
where $p_i>0$ and $\sum_{i=1}^k p_i=1$, $|\psi_i \rangle$ is a pure state and $\langle\psi_i |$ is the complex conjugate transpose of $|\psi_i \rangle$. Hence, the density matrix of $\rho$ is also corresponding to a Hermitian tensor $\mathcal{H}_{\rho}\in \mathbb{H} {[n_1, \ldots, n_m]} $
\begin{equation}\label{eq:Htensmixst}
  \mathcal{H}_{\rho}:=\sum_{i=1}^k p_i \chi_{|\psi_i \rangle}\otimes \chi_{|\psi_i \rangle}^{*},
\end{equation}
where $\chi_{|\psi_i \rangle} $ is the corresponding complex tensor of the state $|\psi_i \rangle $.
A quantum mixed state $\rho$ is called separable if it can be written as
\begin{equation}\label{eq:stateDcomp}
  \rho=\sum_{i=1}^r \lambda_i |\phi_i^{(1)}\cdots \phi_i^{(m)}\rangle \langle \phi_i^{(1)}\cdots \phi_i^{(m)}|,
\end{equation}
where $|\phi_i^{(k)}\rangle$ is the pure state of the $k$-th system $\mathbb{C}^{n_k}$.
Hence, a quantum mixed state $\rho$ is separable if and only if its corresponding tensor $\mathcal{H}_\rho$ has a positive Hermitian decomposition as (\ref{Eq:rank-oneHermitDecomp}).

Given a symmetric Hermitian tensor $\mathcal{S}\in s\mathbb{H} {[n, m]}$.
if $\mathcal{S}$ can be written as
\begin{equation} \label{Eq:posiSHermitDecomp}
\mathcal{S}=\sum_{i=1}^r \lambda_i \,  u_i^{\otimes m} \otimes (u_i^*)^{\otimes m}
\end{equation}
with $0<\lambda_i\in \mathbb{R}$, $u_i\in \mathbb{C}^{n}$
and $\|u_i \|=1$ for all $i\in [r]$,
then (\ref{Eq:posiSHermitDecomp}) is called a {\it symmetric positive Hermitian decomposition} of $\mathcal{S}$.

It is well known that every symmetric complex tensor has a symmetric rank-one decomposition (cf. P. Comon, et al., \cite[Lemma 4.2]{CGL2008}). Numerical examples show that some of positive Hermitian decomposable and symmetric Hermitian tensors also have symmetric positive Hermitian decompositions. Hence, the definition of symmetric positive Hermitian decomposition is not vacuous.

{\it E-truncated K-moment problem.}---The $E$-truncated $K$-moment problems appear frequently in applications, such as sparse polynomial optimization (cf. Lasserre \cite{Las2006}), completely positive matrices decomposition \cite{ZF2014}, symmetric tensor decomposition \cite{Nie2017}.
Let's briefly review some results on the truncated moment problem (TMP). We refer to \cite{Tch1957,Nie2012,FN2012,Nie2014,Nie2017,CF2005} for details about TMP.

Let $E\subseteq \mathbb{N}^n$ be a finite set, $\mathbb{R}^E$ be the space of real vectors indexed by elements in $E$. An {\it $E$-truncated moment sequence} ($E$-tms) is a vector $y=(y_\alpha)_{\alpha\in E}\in \mathbb{R}^E$.
Let $K$ be the semialgebraic set\\
\centerline{$ K := \{x \in \mathbb{R}^n : h(x) = 0, g(x) \geq 0\}$}\\
where $h = (h_1, \cdots , h_{m_1})$ and $g = (g_1, \cdots , g_{m_2})$.
A nonnegative Borel measure $\mu$ on $\mathbb{R}^n$ is called a $K$-measure if its support, denoted by
supp($\mu$), is contained in $K$.
An $E$-tms $y$ is said to admit a $K$-measure $\mu $ if\\
\centerline{$ y_\alpha = \int_K x^\alpha {\rm d}\mu,\ \ {\rm for\ all}\ \alpha\in E.$}\\
The measure $\mu$ satisfying the above is called a {\it K-representing measure} for $y$.
Denote meas($y,K$) as the set of all $K$-measures admitted by $y$.
The $E$-truncated $K$-moment problem (ETKMP) is to determine a given
$E$-tms $y$ whether admits a $K$-measure or not. If it does, how can we get a K-representing measure?

For each tms $z\in \mathbb{R}^{\mathbb{N}^n_d}$, define a {\it Riesz functional} $L_z$ acting on $\mathbb{R}[x]_d$  as
$ L_z ( p ):=\sum_{\alpha\in \mathbb{N}^n_d} p_\alpha z_\alpha, $
where $p=\sum_{\alpha\in \mathbb{N}^n_d} p_\alpha x^\alpha \in \mathbb{R}[x]_d $. We also denote $\langle p, z\rangle := L_z(p) $ for convenient.

For a tms $z\in \mathbb{R}^{\mathbb{N}^n_{2k}}$, define $M_k(z)$ to be the symmetric matrix, which is linear in $z$, such that
$ L_z(p^2) = {\rm vec}(p)^T M_k(z) {\rm vec}(p), {\rm\ for\ all\ } p\in \mathbb{R}[x]_k, $
where ${\rm vec}(p)$ denotes the coefficient vector of $p$ in the graded lexicographical ordering. The matrix $M_k(z)$ is called a $k$-th {\it order moment matrix}.

For $z \in \mathbb{R}^{\mathbb{N}^n_{2k}}$ and $h\in \mathbb{R}[x]_{2k}$, define the $k$-th {\it localizing matrix} of $h$ generated by $z$, denoted by $L^{(k)}_h (z)$, to be the symmetric matrix, which is linear in $z$, such that
$L_z(h p^2) = {\rm vec} (p)^T (L^{(k)}_h (z)) {\rm vec} (p),\ {\rm for\ all\ } p\in \mathbb{R}[x]_{k-\lceil {\rm deg}(h)/2\rceil}. $
As shown by Nie in \cite{Nie2014}, if $z \in \mathbb{R}^{\mathbb{N}^n_{2k}}$ admits a $K$-measure, then
\begin{equation}\label{eq:Lmatrix}
    L^{(k)}_{h_i}(z) = 0, \ \  M_k(z)\succeq 0, \ \ L^{(k)}_{g_j} (z) \succeq 0
\end{equation}
for all $i \in [m_1]$ and $j \in [m_2]$. If $z$ also satisfies the rank condition
\begin{equation}\label{eq:rankMk}
    {\rm rank}\ M_{k-1}(z) = {\rm rank}\ M_k(z),
\end{equation}
then $z$ admits a unique measure, which is $r$-atomic with $r = {\rm rank}\ M_k(z)$. $z$ is called {\it flat} if both (\ref{eq:Lmatrix}) and (\ref{eq:rankMk}) are satisfied.

{\it E-truncated K-moment method for positive Hermitian decomposition.}--
Assume that $\rho$ is an $m$-partite mixed state, $\mathcal{H}\in \mathbb{H}[n_1,\cdots,n_m]$ is a corresponding Hermitian tensor of $\rho$. If $\rho$ is a symmetric mixed state and we will check whether $\rho$ is symmetric separable or not, then we denote its corresponding Hermitian tensor as $\mathcal{S}\in s\mathbb{H}[m,n]$.
Here we consider two kinds of decompositions: Case 1 denotes the positive Hermitian decomposition of $\mathcal{H}$ and case 2 denotes symmetric positive Hermitian decomposition of $\mathcal{S}$.

Case 1: We give some basic notations as
$n:=\sum_{k=1}^m n_k,$ $ L(i):=2\sum_{k=1}^i n_k,$
$ x:=(x^{(1)}, \cdots, x^{(m)})^T\in \mathbb{R}^{2n},$
$ x^{(i)}:=(x_{Re}^{(i)}, x_{Im}^{(i)} )^T\in \mathbb{R}^{2n_i},$
$ x_{Re}^{(i)}:=(x_{L(i-1)+1}, \cdots, x_{L(i-1)+n_i}),$
$ x_{Im}^{(i)}:=(x_{L(i-1)+n_i+1}, \cdots, x_{L(i-1)+2n_i}),$
$ h_i (x^{(i)}):= (x^{(i)})^T x^{(i)} -1,\  i\in [m],$
$ h(x): =(h_1(x^{(1)}),\cdots, h_m(x^{(m)})).$

Define a semialgebraic set
\begin{equation}\label{Eq:CondK1}
  K:=\{x| h(x)=0\}.
\end{equation}
Denote by $\mathfrak{B}(K)$ the set of all atomic Borel $K$-measures.
Assume that a measure $\mu\in\mathfrak{B}(K)$ is defined as
\begin{equation}\label{Eq:BorelMeas}
   \mu:=\sum_{i=1}^r \lambda_i \ \delta_{x|_{[i]}},\ \lambda_i>0,\ x|_{[i]}\in K,\  {\rm for\ all\ } i\in [r].
\end{equation}
Denote $u^{(k)}:=x_{Re}^{(k)}+\sqrt{-1}x_{Im}^{(k)},\ k=1,2, \cdots, m$. Then one can get a tuple $(u^{(1)}_i, u^{(2)}_i, \cdots, u^{(m)}_i) $ for each $x|_{[i]}$ $(i\in [r])$. If the condition
\begin{equation}\label{Eq:H_intDecom}
  \mathcal{H}=\int_K u^{(1)}\otimes \ldots \otimes u^{(m)} \otimes u^{(1)*}\otimes \ldots \otimes u^{(m)*} {\rm d}\mu
\end{equation}
is satisfied, then we obtain a positive Hermitian decomposition of $\mathcal{H}$ as in (\ref{Eq:rank-oneHermitDecomp}). Conversely, for every $\mu\in\mathfrak{B}(K)$ satisfying (\ref{Eq:H_intDecom}), we can always get a positive Hermitian decomposition of $\mathcal{H}$. In turn, it is true. 

Denote $I:=(i_1\cdots i_m)$, $J:=(j_1\cdots j_m)$, $E_\mathcal{H}:=\{(i_1,\cdots, i_m)|i_k\in [n_k], k\in [m] \}$ as the half-subscript set of $\mathcal{H}$. Let $P_{IJ}(x):= \prod_{k=1}^m (u^{(k)})_{i_k}\ (u^{(k)*})_{j_k}. $
Then the condition (\ref{Eq:H_intDecom}) can be rewritten as \\
\centerline{(CD1):\ \ \ $ \mathcal{H}_{IJ}=\int_K P_{IJ}(x) {\rm d}\mu,\ {\rm for\ all\ } I, J\in E_\mathcal{H}. $}

Case 2: Let $\mathcal{S}\in s\mathbb{H}[m,n]$. We also give some basic notations as
$ x:=(x_{Re}, x_{Im} )^T\in \mathbb{R}^{2n},$
$ x_{Re}:=(x_1, \cdots, x_{n}),$ $ x_{Im}:=(x_{n+1}, \cdots, x_{2n}),$
$ h_1 (x):= x^T x -1,$ $ h(x):=(h_1(x)).$
Assume that $K$ is defined as in (\ref{Eq:CondK1}), $\mathfrak{B}(K)$ is the set of all atomic Borel $K$-measures, and $\mu\in\mathfrak{B}(K)$ as in (\ref{Eq:BorelMeas}).
Let $u=x_{Re}+\sqrt{-1}x_{Im}$. If the condition
\begin{equation}\label{Eq:H_intDecomsym}
  \mathcal{S}=\int_K u^{\otimes m} \otimes (u^*)^{\otimes m} {\rm d}\mu
\end{equation}
is satisfied, then we have a symmetric positive Hermitian decomposition of $\mathcal{S}$ as in (\ref{Eq:posiSHermitDecomp}).
Let $P_{IJ}(x):= \prod_{k=1}^m u_{i_k}\  u_{j_k}^* .$ Then (\ref{Eq:H_intDecomsym}) can be reformulated as\\
\centerline{(CD2):\ \ \ $ \mathcal{S}_{IJ}=\int_K P_{IJ}(x) {\rm d}\mu,\ {\rm \ for\ all\ } I, J\in E_\mathcal{S}. $}\\
From the above discussion, we get the following theorem about the separability discrimination of mixed states without proof.

\begin{theorem}\label{Th:intPHDcond}\rm
(i) Assume that $\rho$ is an $m$-partite quantum mixed state, $\mathcal{H}\in \mathbb{H}[n_1,\cdots, n_m]$ is a corresponding Hermitian tensor of $\rho$ and $K$ is defined as in (\ref{Eq:CondK1}). Then  $\rho$ is separable if and only if there exists a measure $\mu\in \mathfrak{B}(K)$ such that condition (CD1) is satisfied.

(ii) Assume that $\rho$ is an $m$-partite symmetric mixed state, and $\mathcal{S}\in s\mathbb{H}[m,n]$ is a corresponding Hermitian tensor of $\rho$. Then
$\rho$ is symmetric separable if and only if there exists a measure $\mu\in \mathfrak{B}(K)$ such that condition (CD2) is satisfied.
\end{theorem}


{\it A hierarchy of semidefinite relaxation.}---Denote $R_{IJ}(x)+\sqrt{-1}T_{IJ}(x):=P_{IJ}(x)$,
 i.e., $R_{IJ}(x)$ and $T_{IJ}(x)$ are the real part and the image part of $P_{IJ}(x)$ respectively.
Choose a generic sum of square (SOS) polynomial $F(x)$. Consider the linear optimization
\begin{eqnarray}
\label{min:HintLiner}  \min_\mu &\ & \int_K F(x) {\rm d}\mu 
\end{eqnarray}

\begin{center}\begin{tabular}{c l}
  s.t. & ${\rm Re}\ \mathcal{H}_{IJ}=\int_K R_{IJ}(x) {\rm d}\mu, \ (I, J\in E_\mathcal{H})$ \\
       & $ {\rm Im}\ \mathcal{H}_{IJ}=\int_K T_{IJ}(x) {\rm d}\mu,\ (I, J\in E_\mathcal{H})$\\
       & $ \mu\in \mathfrak{B}(K),$ \\
   or  & $ {\rm Re}\ \mathcal{S}_{IJ}=\int_K R_{IJ}(x) {\rm d}\mu, \ (I, J\in E_\mathcal{S})$\\
       & $ {\rm Im}\ \mathcal{S}_{IJ}=\int_K T_{IJ}(x) {\rm d}\mu,\ (I, J\in E_\mathcal{S})$\\
       & $ \mu\in \mathfrak{B}(K).$
\end{tabular}\end{center}

{\bf Remark 1.} (I) Assume that $\rho$ is an $m$-partite mixed state, $\mathcal{H}$ is a corresponding Hermitian tensor of $\rho$. If the first part of the constraint condition of (\ref{min:HintLiner}), i.e.  the constraint condition on $\mathcal{H}$, is infeasible then the state $\rho$ is not separable.
(II) When $\rho$ is an $m$-partite symmetric mixed state and we consider its symmetric separability, we denote $\mathcal{S}\in s\mathbb{H}[m,n]$ as its corresponding Hermitian tensor. If the second part of the constraint condition of (\ref{min:HintLiner}) is infeasible then $\rho$ is not symmetric separable.

In order to solve (\ref{min:HintLiner}), we replace $\mu$ by the vector of its moments.
Let $d > 2m$ be an even integer and $d\leq 2k$. Denote the moment cones\\
$\mathfrak{C}_d = \left\{y\in \mathbb{R}^{\mathbb{N}^{2n}_d} |  y_\alpha =\int_K x^\alpha {\rm d}\mu, \  \alpha\in \mathbb{N}^{2n}_d,\ \mu\in \mathfrak{B}(K)   \right\},$\\
$\mathfrak{C}^k = \left\{y\in \mathbb{R}^{\mathbb{N}^{2n}_{2k}} |  M_k(y)\succeq 0,\ L_{h_i}^k(y)=0,\  h_i\in h   \right\},$ \\
$\mathfrak{C}_d^k = \left\{y\in \mathbb{R}^{\mathbb{N}^{2n}_{d}} | \exists z\in \mathfrak{C}^k,\ y=z|_d  \right\}.$\\
It is clear that, for all $k\geq d/2$,\\
\centerline{$
\mathfrak{C}_d \subseteq \mathfrak{C}_d^{k+1}\subseteq \mathfrak{C}_d^{k}\ {\rm and}\ \mathfrak{C}_d=\bigcap_{k\geq d/2} \mathfrak{C}_d^{k}.
$}
Then (\ref{min:HintLiner}) is equivalent to the linear optimization\\
\begin{eqnarray}
\label{min:HmomLiner}  \min_y &\ & \langle F, y\rangle \\
\nonumber   s.t. &\ &  {\rm Re}\ \mathcal{H}_{IJ}=\langle R_{IJ}, y\rangle , \ (I, J\in E_\mathcal{H})\\
\nonumber        &\ & {\rm Im}\ \mathcal{H}_{IJ}=\langle T_{IJ},y \rangle ,\ (I, J\in E_\mathcal{H})\\
\nonumber   {\rm or} &\ &  {\rm Re}\ \mathcal{S}_{IJ}=\langle R_{IJ}, y\rangle , \ (I, J\in E_\mathcal{S})\\
\nonumber        &\ & {\rm Im}\ \mathcal{S}_{IJ}=\langle T_{IJ},y \rangle ,\ (I, J\in E_\mathcal{S})\\
\nonumber        &\ & y\in \mathfrak{C}_d.
\end{eqnarray}


By \cite[Curto and Fialkfalt]{CF2005}, $y$ admits a unique K-measure, which is rank$M_k(y)$-atomic, if $y$ is flat. 
And the cone $\mathfrak{C}_d$ can be approximated by semidefnite relaxations.
This follows the hierarchy of semidefinite relaxations
\begin{eqnarray}
\label{min:HmLSDR}  \min_\mu &\ & \langle F, y\rangle \\
\nonumber   s.t. &\ &  {\rm Re}\ \mathcal{H}_{IJ}=\langle R_{IJ}, y\rangle , \ (I, J\in E_\mathcal{H})\\
\nonumber        &\ & {\rm Im}\ \mathcal{H}_{IJ}=\langle T_{IJ},y \rangle ,\ (I, J\in E_\mathcal{H})\\
\nonumber   {\rm or} &\ &  {\rm Re}\ \mathcal{S}_{IJ}=\langle R_{IJ}, y\rangle , \ (I, J\in E_\mathcal{S})\\
\nonumber        &\ & {\rm Im}\ \mathcal{S}_{IJ}=\langle T_{IJ},y \rangle ,\ (I, J\in E_\mathcal{S})\\
\nonumber        &\ & y\in \mathfrak{C}^k,
\end{eqnarray}
for $k=d/2,\ d/2+1, \cdots$. The following algorithm is applied to solve the hierarchy of (\ref{min:HmLSDR}). It can check whether a (symmetric) mixed state is (symmetric) separable or not and give a (symmetric) decomposition for it if it is.

\begin{algo}\label{Al:ETKMSDR}\rm
{(\bf ETKM-SDR method for (symmetric) positive Hermitian decomposition).}

{\bf Input}: A tensor $\mathcal{H}\in \mathbb{H}[n_1,\cdots,n_m]$ or $\mathcal{S}\in s\mathbb{H}[m,n]$.

{\bf Output}: Whether the tensor has a (symmetric) positive Hermitian decomposition or not and giving a (symmetric) positive Hermitian decomposition for it if it has.

{\bf Step 1}:  Set $d=2(m+1)$. Choose a generic SOS function $F(x)$ of degree at most $d$. Let $k=d/2$.

{\bf Step 2}: Solve (\ref{min:HmLSDR}). If (\ref{min:HmLSDR}) is infeasible, then $\mathcal{H}$ (or $\mathcal{S}$) has not a (symmetric) positive Hermitian decomposition, and stop. Otherwise, compute a minimizer $y^k$. Let $t := 1$.

{\bf Step 3}: Let $z=y^k|_{2t}$. If the rank condition (\ref{eq:rankMk}) is satisfied, go to Step 5.

{\bf Step 4}: If $t < k$, set $t= t + 1$ and go to Step 3; otherwise, set $k= k + 1$ and
go to Step 2.

{\bf Step 5}: Compute $r={\rm rank}M_t(z)$, $\lambda_1, \cdots, \lambda_r>0$, $x|_{[1]}, \cdots, x|_{[r]}\in K$. Output a (symmetric) positive Hermitian decomposition of $\mathcal{H}$ (or $\mathcal{S}$) as (\ref{Eq:rank-oneHermitDecomp}) (or (\ref{Eq:posiSHermitDecomp})).

\end{algo}

\begin{theorem}\label{Th:algo1}\rm
Algorithm \ref{Al:ETKMSDR} has the following properties: (I) If (\ref{min:HmLSDR}) is infeasible for some k, then $\mathcal{H}$ (or $\mathcal{S}$) has not a (symmetric) positive Hermitian decomposition, i.e., $\rho$ is not (symmetric) separable;
(II) If $\mathcal{H}$ (or $\mathcal{S}$) has a (symmetric) positive Hermitian decomposition, i.e., $\rho$ is (symmetric) separable, then for almost all generated $F(x)$, we can asymptotically get a (symmetric) positive Hermitian decomposition by solving the hierarchy of semidefinite relaxations (\ref{min:HmLSDR}) for $k$ big enough.
\end{theorem}

{\it Proof.} The proof is in the Appendix.

{\bf Remark 2.} In calculating, we choose subscripts $(I,J)$ in (\ref{min:HmLSDR}) meeting the following requirements.

(I) If $\mathcal{H}$ is an Hermitian tensor, then $
\mathcal{H}_{IJ}=\int_K P_{IJ}(x) {\rm d}\mu$ if and only if $
H_{JI}=\int_K P_{JI}(x) {\rm d}\mu$. Hence we may choose subscripts $(I,J)$ in (\ref{min:HmLSDR}) satisfying $i_1<j_1$, or $ i_1=j_1$ but $i_2<j_2$, and so on, till $i_k=j_k$ for all $k\in [m-1]$ but $i_m \leq j_m$, which is denoted by $I\leq J$.

(II) If $\mathcal{S}$ is an Hermitian and symmetric tensor, then we choose subscripts $(I,J)$ in (\ref{min:HmLSDR}) satisfying $I\leq J$ and $1\leq i_1\leq i_2\leq\cdots\leq i_m\leq n$.

{\it Numerical examples.}---Given a quantum mixed state $\rho$. We will use Algorithm \ref{Al:ETKMSDR} to check its separability and give a decomposition if it is separable.
In the processing, we first rewrite a state as its corresponding Hermitian tensor, then check the Hermitian tensor whether has a (symmetric) positive Hermitian decomposition and get a decomposition if it has, finally write the decomposition as the states form.
We use the toolbox \emph{Gloptipoly 3} \cite{HLL2009} and \emph{SeDuMi} \cite{SDM1999} to solve the SDR problems (\ref{min:HmLSDR}).

\begin{example}\rm\label{Ex: 3qubit}
(3-qubit system) Consider a mixed state
$\rho(\frac{1}{4},\frac{3}{8})=\frac{1}{4}|GHZ\rangle\langle GHZ|+\frac{3}{8}|W\rangle\langle W|+\frac{3}{8}|\tilde{W}\rangle\langle \tilde{W}|$ proposed by Wei and Goldbart in \cite{Wei2003}. Here, $|GHZ\rangle$, $|W\rangle$ and $\tilde{W}\rangle$ are defined as
$|GHZ\rangle=(|000\rangle+|111\rangle)/\sqrt{2}$, $|W\rangle=(|001\rangle+|010\rangle+|100\rangle)/\sqrt{3}$, $|\tilde{W}\rangle=(|110\rangle+|101\rangle+|011\rangle)/\sqrt{3}.$
Wei and Goldbart said that $\rho(\frac{1}{4},\frac{3}{8})$ is separable.
But they did not give its decomposition.

We first write $|GHZ\rangle$, $|W\rangle$ and $|\tilde{W}\rangle$ corresponding tensors $\chi_{|GHZ\rangle}$, $\chi_{|W\rangle}$ and $\chi_{|\tilde{W}\rangle}$, respectively. Their nonzero entries are $(\chi_{|GHZ\rangle})_{111}=(\chi_{|GHZ\rangle})_{222}=\frac{1}{\sqrt{2}}$,
$(\chi_{|W\rangle})_{112}=(\chi_{|W\rangle})_{121}
=(\chi_{|W\rangle})_{211}=\frac{1}{\sqrt{3}}$,
$(\chi_{|\tilde{W}\rangle})_{122}=(\chi_{|\tilde{W}\rangle})_{212}
=(\chi_{|\tilde{W}\rangle})_{221}=\frac{1}{\sqrt{3}}$.
Hence, the corresponding Hermitian tensor of $\rho(\frac{1}{4},\frac{3}{8})$ is given by
$\mathcal{H}=
\frac{1}{4}\chi_{|GHZ\rangle}\otimes \chi_{|GHZ\rangle}^{*}
+\frac{3}{8}\chi_{|W\rangle}\otimes \chi_{|W\rangle}^{*}
+\frac{3}{8}\chi_{|\tilde{W}\rangle}\otimes \chi_{|\tilde{W}\rangle}^{*}.$


Obviously, $\mathcal{H}$ is symmetric. Hence, we attempt to decompose $\mathcal{H}$ with the symmetric case. Fortunately, we succeeded. Applying Algorithm \ref{Al:ETKMSDR}, we get a symmetric positive Hermitian decomposition as
$\mathcal{H}=\frac{1}{3}u_1^{\otimes 3}\otimes u_1^{*\otimes 3}
+\frac{1}{3}u_2^{\otimes 3}\otimes u_2^{*\otimes 3}
+\frac{1}{3}u_3^{\otimes 3}\otimes u_3^{*\otimes 3},$
where $u_1=(0.1222-0.6965i,0.1222-0.6965i)^{T}$,$u_2=(0.5293+0.4689i,0.1414-0.6928i)^{T}$,
$u_3=(-0.4830-0.5165i,0.6888-0.1601i)^{T}$.
Let\\
 $|\phi_1\rangle=(0.1222-0.6965i)|0\rangle+(0.1222-0.6965i)|1\rangle$,
 $ |\phi_2\rangle= (0.5293+0.4689i)|0\rangle+(0.1414-0.6928i)|1\rangle$,
 $ |\phi_3\rangle= (-0.4830-0.5165i)|0\rangle+(0.6888-0.1601i)|1\rangle$.
Then we get a symmetric decomposition of $\rho(\frac{1}{4},\frac{3}{8})$ as\\  $  \rho=\frac{1}{3}|\phi_1\rangle^{\otimes 3}\langle \phi_1|^{\otimes 3}
   +\frac{1}{3}|\phi_2\rangle^{\otimes 3}\langle \phi_2|^{\otimes 3}
+\frac{1}{3}|\phi_3\rangle^{\otimes 3}\langle \phi_3|^{\otimes 3}. $ \qed

\end{example}

\begin{example}\rm\label{Ex: 2qubit}
(Two-qubit system) We consider the following the bipartite qubit mixed state (cf. Hu et al. \cite[Example 1]{Hu2012})
\begin{eqnarray*}
  \rho&=&\frac{1}{2}\left( \frac{1}{\sqrt{2}}|00\rangle+|11\rangle\right)
\left( \frac{1}{\sqrt{2}}\langle00|+\langle11|\right) \\
  && +\frac{1}{2}\left( \frac{1}{\sqrt{2}}|01\rangle+|10\rangle\right)
\left( \frac{1}{\sqrt{2}}\langle01|+\langle10|\right).
\end{eqnarray*}
Hu et al. computed that the geometric measure of $\rho$ was equal to zero. Hence, the state is separable. And they gave a non-symmetric decomposition with four terms.
So we attempt to decompose ${\rho}$ by Algorithm \ref{Al:ETKMSDR} with the symmetric case and obtain a symmetric decomposition with two terms as
$\rho=\frac{1}{2}|\phi_1\phi_1\rangle\langle \phi_1\phi_1|
+\frac{1}{2}|\phi_2\phi_2\rangle\langle \phi_2\phi_2|,$ where
$|\phi_1\rangle =(-0.5992-0.3754i)|0\rangle-(0.5992+0.3754i)|1\rangle $, $|\phi_2\rangle=(-0.4303+0.5611i)|0\rangle+(0.4303-0.5611i)|1\rangle $. \qed

\end{example}

{\bf Remark 2.} From Example \ref{Ex: 3qubit} and \ref{Ex: 2qubit}, we find that we may obtain a symmetric decomposition when the given mixed state is separable and symmetric. A natural question: Does a  symmetric and separable mixed state always has a symmetric decomposition?
However, the following example tells us that it doesn't happen all the time.

\begin{example}\rm\label{Ex:sHpsdNsym}
($m$-partite $n$-dimension system) Let $|\phi_i\rangle\in \mathbb{C}^n $ ( $i\in [m]$) be nonzero normalized states and different from each other. Denote by $per(m)$ as the set of all permutations of $\{1,2,\cdots,m\}$. Let\\
\centerline{$ \rho=\frac{1}{m!}\sum_{P\in per(m)} |\phi_{P(i_1)}\cdots\phi_{P(i_m)}\rangle \langle \phi_{P(i_1)}\cdots\phi_{P(i_m)}|.$}
It is clear that the state $\rho$ is an $m$-partite $n$-dimension symmetric mixed state and has a non-symmetric decomposition. However, applying Algorithm \ref{Al:ETKMSDR}, we find that these states  have not their symmetric decompositions for $m=2,\ n=2, 3, 4$ and $m=3,\ n=2, 3, 4$. All $|\phi_i\rangle $ ( $i\in [m]$) are obtained random in the numerical example. Hence, we guess that this kind of symmetric mixed state has no symmetric decomposition. \qed
\end{example}

\begin{example}\rm\label{Ex:shortenlength}
(Random) Let $\rho$ be a symmetric separable $m$-partite $n$-dimension mixed state. It is formulated as\\
\centerline{
   $\rho=(1/r)\sum_{k=1}^r |\phi_k\rangle^{\otimes m} \langle\phi_k|^{\otimes m}.$}\\
Take $m=2,\ n=2,\ r=7$. $|\phi_k\rangle$ are obtained randomly\\
 $ |\phi_1\rangle = (0.1865-0.0210 i)|0 \rangle  + (0.7198+0.6684 i) |1 \rangle $,  \\
 $ |\phi_2\rangle = (0.3857+0.5437 i)|0 \rangle  + ( 0.4330+0.6067 i) |1 \rangle $,  \\
 $ |\phi_3\rangle = (0.5296+0.6557 i)|0 \rangle  + (0.3881+0.3800 i) |1 \rangle $,  \\
 $ |\phi_4\rangle = (-0.0967+0.8886 i)|0 \rangle  + (0.0243+0.4477 i) |1 \rangle $,  \\
 $ |\phi_5\rangle = (0.7262+0.4409 i)|0 \rangle  + (0.1067+0.5166 i) |1 \rangle $,  \\
 $ |\phi_6\rangle = (-0.1165+0.4494 i)|0 \rangle  + ( 0.8815-0.0864 i) |1 \rangle $,  \\
 $ |\phi_7\rangle = (0.9394+0.0557  i)|0 \rangle  + ( 0.2231+0.2544 i) |1 \rangle $.\\
By Algorithm \ref{Al:ETKMSDR}, we obtain a rank-4 decomposition $ \rho=\sum_{k=1}^4 p_k |\Phi_k\rangle^{\otimes 2} \langle\Phi_k|^{\otimes 2}$, where
$p_1=0.5239,\ p_2=0.1430,\ p_3=0.1978,\ p_4=0.1354$, and\\
$ |\Phi_1\rangle =  (-0.6441-0.6383 i) |0\rangle  + (-0.1795-0.3814 i) |1\rangle, $ \\
$ |\Phi_2\rangle =  (-0.1910-0.2962 i) |0\rangle  + (0.5569-0.7521 i) |1\rangle, $ \\
$ |\Phi_3\rangle =  (0.5179-0.2372 i) |0\rangle  + (0.5682-0.5939 i) |1\rangle, $ \\
$ |\Phi_4\rangle =  (0.1643+0.4561 i) |0\rangle  + (0.5948-0.6412 i) |1\rangle $. \qed

{\bf Remark 3.} Example \ref{Ex:shortenlength} tells us that if a mixed state $\rho$ is separable and given by a sum of many terms, then we may reduce the number of terms by Algorithm \ref{Al:ETKMSDR}.

\begin{example}\rm
(Non-symmetric decomposition) Consider a bipartite $n$-dimensional mixed state\\
\centerline{$\rho_{iso}(F)=\frac{1-F}{n^2-1}\left(\mathbb{I}-|\Phi^{+}\rangle\langle\Phi^{+}|\right)
+F|\Phi^{+}\rangle\langle\Phi^{+}|$,}\\
where $|\Phi^{+}\rangle=\frac{1}{\sqrt{n}}\sum_{i=1}^{n}|ii\rangle$. $\rho_{iso}(F)$ is called isotropic state in \cite{Wei2003} and known to be separable for $F\in[0,\frac{1}{n}]$ by M. Horodecki and P. Horodecki in  \cite{Horodecki1999}.
Obviously, the state is symmetric. Let $n=2$, $F=\frac{1}{2}$. We find that the state has no symmetric decomposition by Algorithm \ref{Al:ETKMSDR}. But we calculate a non-symmetric decomposition\\
\centerline{$\rho_{iso}(1/2)=\sum_{k=1}^{5}p_k|\phi_k^{(1)}\phi_k^{(2)}\rangle
 \langle \phi_k^{(1)} \phi_k^{(2)}|$,}\\
where $p_1=0.2476,\ p_2=0.2496,\ p_3=0.1257, \ p_4=0.2450,\ p_5=0.1323$, and\\
$|\phi_1^{(1)}\rangle=(0.2008-0.6093i)|0\rangle+(0.4979+0.5834i)|1\rangle,$\\
$|\phi_1^{(2)}\rangle=(-0.1246-0.6294i)|0\rangle-(0.5656-0.5180i)|1\rangle,$\\
$|\phi_2^{(1)}\rangle=(0.8416+0.5326i)|0\rangle+(0.0886-0.0110i)|1\rangle$,\\
$|\phi_2^{(2)}\rangle=(-0.9062-0.4132i)|0\rangle-(0.0393+0.0801i)|1\rangle$,\\
$|\phi_3^{(1)}\rangle=(-0.5960-0.1036i)|0\rangle+(0.4408+0.6631i)|1\rangle$,\\
$|\phi_3^{(2)}\rangle=(0.5325-0.2864i)|0\rangle-(0.2088-0.7686i)|1\rangle$,\\
$|\phi_4^{(1)}\rangle=(0.4495+0.2734i)|0\rangle+(0.6874+0.5007i)|1\rangle$,\\
$|\phi_4^{(2)}\rangle=(-0.5259-0.0137i)|0\rangle-(0.8491-0.0484i)|1\rangle$,\\
$|\phi_5^{(1)}\rangle=(-0.4713+0.2360i)|0\rangle+(0.6299+0.5703i)|1\rangle$,\\
$|\phi_5^{(2)}\rangle=(0.4148-0.3256i)|0\rangle+(0.2467+0.8129i)|1\rangle$. \qed

\end{example}


\end{example}


{\it Conclusions.}---We have presented an approach for checking whether an $m$-partite quantum mixed state is separable or not and give a decomposition for it if it is. Our approach is valid for every $m$-partite mixed state. The approach relies on the $E$-truncated $K$-moment problem and semidefinite relaxations method. Algorithm \ref{Al:ETKMSDR} can be used for symmetric and non-symmetric decomposition of separable mixed states.

By numerical experiments, we can find some properties of mixed states. For examples: (I) some symmetric and separable states have symmetric decompositions, but some do not; (II) if a mixed state $\rho$ can be separable as in (\ref{eq:stateDcomp}), then there exists a decomposition such that the number $r$ is the smallest. We call the smallest $r$ as the rank of $\rho$ or the symmetric rank of $\rho$ if the decomposition is symmetric.
From large number of numerical experiments, we find that the upper bounds of symmetric rank of two-qubit and three-qubit mixed states are 4 and 7, respectively.
We may further study some properties of mixed separable states in the future by the algorithm.

We thank Jinyan Fan and Anwa Zhou for helpful discussions. The research is supported by the National Natural Science Foundation of China (No. 11871472).


{\it Appendix.}---{\bf Proof of Theorem \ref{Th:algo1}}:
(I) We only prove the case $\mathcal{H}$, since the case $\mathcal{S}$ can be deduced similarly. Suppose that $\mathcal{H}$ has a positive Hermitian decomposition as (\ref{Eq:rank-oneHermitDecomp}) with $\lambda_i>0$ for all $i\in [r]$. Let $u_i^{(k)}=x_{Re}^{(k)}|_{[i]}+\sqrt{-1}x_{Im}^{(k)}|_{[i]}$, $ x^{(k)}|_{[i]}=(x_{Re}^{(k)}|_{[i]}, x_{Im}^{(k)}|_{[i]} )^T$, $x|_{[i]}=(x^{(1)}|_{[i]}, \cdots, x^{(m)}|_{[i]})^T $. Then  vectors $x|_{[1]}, \cdots, x|_{[r]}\in K$. Take the weighted Dirac measure
$ \mu=\sum_{i=1}^r \lambda_i \ \delta_{x|_{[i]}}. $  Then, there exists a tms $y\in \mathbb{R}^{\mathbb{N}_d^{2n}}$ admitting the measure $\mu$ such that ${\rm Re}\ \mathcal{H}_{IJ}=\langle R_{IJ}, y\rangle$ and $ {\rm Im}\ \mathcal{H}_{IJ}=\langle T_{IJ}, y\rangle $ for all $I, J\in E_\mathcal{H}$. Furthermore, for all $k\geq d/2$, the tms $z\in \mathfrak{C}^k$ such that $y=z|_{d}$ is feasible for (\ref{min:HmLSDR}), which is a contradiction.

(II) The conclusions can be deduced from Nie \cite[Section 5]{Nie2014}. \ \ $\Box$


\begin{thebibliography}{99}
\bibitem{Ein35}
A. Einstein, B. Podolsky, and N.Rosen, Can quantum-mechanical description of physical reality be
considered completely?  Phys. Rev. {\bf 47}, 777 (1935)

\bibitem{Sch35}
E. Schir\"{o}dinger, Die gegenw$\ddot{a}$rtige Situation in der Quantenmechanik, Naturwissenschaften {\bf 23}, 807-812 (1935)

\bibitem{Guh2004}
O. G\"{u}hne, Characterizing Entanglement via Uncertainty Relations, Phys. Rev. Lett. {\bf 92}, 117903 (2004).

\bibitem{LWFL2014}
M. Li, J. Wang, S-M Fei, and X. Li-Jost, Quantum separability criteria for arbitrary-dimensional multipartite states, Phys. Rev. A {\bf 89}, 022325 (2014)

\bibitem{Bell1964}
J.S. Bell, On the Einstein-Podolsky-Rosen paradox, Physics 1 (1964) 195. Reprinted in J. Bell, Speakable and Unspeakable in Quantum Mechanics, Cambridge University Press, 2004

\bibitem{Per1996}
 A. Peres, Separability criterion for density matrices. Phys. Rev. Lett. {\bf 77}, 1413 (1996)

\bibitem{HHH1996}
 M. Horodecki, P. Horodecki, and R. Horodecki, Separability of mixed states: necessary and sufficient conditions. Phys. Lett. A {\bf 223}, 1 (1996)

\bibitem{ZHS1998}
K. Zyczkowski, P. Horodecki, A. Sanpera, and M. Lewenstein, On the volume of the set of mixed entangled states, Phys. Rev. A {\bf 58} (1998) 883.

\bibitem{VW2002}
G. Vidal, and R.F. Werner, A computable measure of entanglement, Phys. Rev. A {\bf 65} (2002) 032314.

\bibitem{HT2003}
H.F. Hofmann, and S. Takeuchi, Violation of local uncertainty relations as a signature of entanglement, Phys. Rev. A {\bf 68} (2003) 032103.

\bibitem{GHG2007}
O. G\"{u}hne, P. Hyllus, O. Gittsovich, and J. Eisert, Covariance matrices and the separability problem. Phys. Rev. Lett. {\bf 99}, 130504 (2007)

\bibitem{Vic2007}
J. D. Vicente, Separability criteria based on the Bloch representation of density matrices. Quant. Inf. Comput. {\bf 7}, 624 (2007)

\bibitem{Ter2000}
B. Terhal, Bell inequalities and the separability criterion. Phys. Lett. A {\bf 271}, 319 (2000)

\bibitem{LKC2000}
M. Lewenstein, B. Kraus, J.I. Cirac, and P. Horodecki, Optimization of entanglement witnesses. Phys. Rev. A {\bf 62}, 052310 (2000)

\bibitem{OH2010}
A. Osterloh1, and P. Hyllus1, Estimating multipartite entanglement measures, Phys. Rev. A {\bf 81}, 022307 (2010)


\bibitem{Bern2015}
Alex E. Bernardini et al, Entanglement and separability in the
noncommutative phase-space scenario, 2015 J. Phys., Conf. Ser. {\bf 626} 012046 (2015)

\bibitem{XZZ2016}
Y. Xi, Z-J Zheng, and C-J Zhu, Entanglement detection via general SIC-POVMs, Quantum Inf. Process {\bf 15}, 5119-5128 (2016)

\bibitem{SLL2018}
S-Q Shen, M. Li, X. Li-Jost, and S-M Fei, Improved separability criteria via some classes of measurements, Quantum Inf. Process {\bf 17}, 111(2018)

\bibitem{NQB14}
G. Ni, L. Qi, and M. Bai, Geometric measure of entanglement and U-eigenvalues of tensors. SIAM J. Matrix Anal. Appl.  {\bf 35}, 73-87 (2014)

\bibitem{CGL2008}
P. Comon, G. Golub, L-H Lim, and B. Mourrain, Symmetric tensors and symmetric tensor rank, SIAM J. Matrix. Anal. Appl. {\bf 30}, 1254-1279 (2008)

\bibitem{Las2006}
J. B. Lasserre. Convergent SDP-relaxations in polynomial optimization with sparsity. SIAM J. Optim.
{\bf 17}, 822-843 (2006)

\bibitem{ZF2014}
A. W. Zhou, and J. Y. Fan, The CP-matrix completion problem, SIAM J. Matrix Anal. Appl. {\bf 35}, 127-142 (2014)

\bibitem{Nie2017}
J. Nie, Generating polynomials and symmetric tensor decompositions, Found. Comput. Math. {\bf 17}, 423-465 (2017)

\bibitem{Tch1957}
V. Tchakaloff, Formules de cubatures m$\acute{e}$canique $\grave{a}$ coefficients non n$\acute{e}$gatifs. Bull. Sci. Math. {\bf 81}, 123-134 (1957)

\bibitem{Nie2012}
J. Nie, A semidefinite approach for truncated $K$-moment problems, Found. Comput. Math. {\bf 12}, 851-881 (2012)

\bibitem{FN2012}
L. Fialkow, and J. Nie. The truncated moment problem via homogenization and flat extensions. J.
Functional Analysis {\bf 263}, 1682-1700 (2012)

\bibitem{Nie2014}
J. Nie, The $A$-Truncated $K$-Moment Problem, Found. Comput. Math. {\bf 14}, 1243-1276 (2014)

\bibitem{CF2005}
R. Curto, and L. Fialkow. Truncated K-moment problems in several variables. J. Operator
Theory {\bf 54}, 189-226 (2005)

\bibitem{HLL2009}
D. Henrion, J. Lasserre, and J. Loefberg, GloptiPoly 3: moments, optimization and
semidefinite programming, Optim. Methods Softw. {\bf 24}, 761-779 (2009)

\bibitem{SDM1999}
J.F. Sturm, SeDuMi 1.02: AMATLAB toolbox for optimization over symmetric cones. Optim. Methods
Softw. {\bf 11 $\&$ 12}, 625-653 (1999)

\bibitem{Wei2003}
T-C Wei, and Paul M. Goldbart, Geometric measure of entanglement and applications to bipartite and multipartite quantum states, Phys. Rev. A {\bf 68}, 042307 (2003).

\bibitem{Hu2012}
S. Hu, L. Qi, Y. Song, and G. Zhang, Geometric measure of entanglement of multipartite mixed states,
Int. J. Software Informatics {\bf 8}, 317-326 (2014)

\bibitem{Horodecki1999}
M. Horodecki, and P. Horodecki, Reduction criterion of separability and limits for a class of distillation protocols, Phys. Rev. A {\bf 59}, 4206-4216(1999).


\end{thebibliography}

\end{document}